\documentclass[prl,letterpaper,twocolumn,showpacs,superscriptaddress]{revtex4}

\usepackage{graphicx,psfrag,amsmath,amssymb,amsfonts,bbm,latexsym,color,dcolumn,bm}

\begin{document}

\title{Neutrino mass variability due to nonminimal coupling to spacetime curvature 
in neutrinophilic two-Higgs-doublet models}

\author{Roberto Onofrio}

\affiliation{Dipartimento di Fisica e Astronomia ``Galileo Galilei'', 
Universit\`a  di Padova, Via Marzolo 8, Padova 35131, Italy}

\affiliation{ITAMP, Harvard-Smithsonian Center for Astrophysics, 
60 Garden Street, Cambridge, MA 02138, USA}
\date{\today}

\begin{abstract}
In neutrinophilic two-Higgs-doublet models, neutrinos acquire mass due to 
a Higgs field with vacuum expectation value of the order of $\simeq 10^{-2}$ eV, 
corresponding to a Compton wavelength in the 10 $\mu$m range.
This creates a situation in which nonminimal couplings between Higgs fields and 
spacetime curvature may lead to novel observable effects. 
Among these, we discuss the possibility of variable neutrino masses, with 
implications for the dependence of the neutrino oscillation frequency on the 
spacetime curvature, a further source of dispersion of the neutrino arrival 
times from supernovae events, and possibly also a mechanism leading to 
gravitationally-induced neutrino superluminality. Finally, we propose 
laboratory-scale experiments in which properly designed electroweak cavities 
may be used to change neutrino masses, which should be observable through time 
of flight experiments.
\end{abstract}

\pacs{04.62.+v, 14.60.Pq, 14.80.Bn}

\maketitle

The mass of elementary particles is, in the standard model, a dynamical observable depending 
on the vacuum expectation value (VEV) of the Higgs field and the Yukawa couplings.
As such, it is possible in principle to change the mass by deforming the VEV of the Higgs 
field through macroscopic external fields. Since the Higgs has no coupling to the photons 
at tree level, this leaves open only the possibility of a nonminimal coupling to a strong-curvature 
region of spacetime. Unfortunately, due to the large mismatching between the typical space-time 
curvature lenghtscale and the Compton wavelength of the Higgs field, this possibility becomes 
realistic only in the presence of very large nonminimal couplings.  This has been discussed in 
\cite{Onofrio}, leading to the prediction of Higgs-related spectroscopic shifts due to changes 
of the electron mass in proximity of strong-gravity sources. The experimental observability of 
mass shifts at moderate values of the nonminimal coupling becomes more favourable if, as in 
various extensions of the standard model involving multiple Higgs doublets, some VEVs are 
intrinsically smaller. This occurs for instance in the context of the so-called neutrinophilic 
two-Higgs-doublet models \cite{Wang,Gabriel}, in which neutrinos acquire mass through coupling 
to a Higgs doublet with VEV corresponding to a Compton wavelength of macroscopic size. 

In this note, we discuss possible implications of such a Higgs doublet on the physics and 
astronomy of neutrinos. More in general, we want to focus the attention on a possibility that 
has not yet been explored to the best of our knowledge, {\it i.e.} that neutrinos, due to their 
intrinsically minute masses, are sensitive probes of the interplay between the curvature of 
spacetime and the Lagrangian of the standard model or its extensions. Then the following 
discussion has to be considered as a specific example of a more general framework, pioneered 
by Stodolsky \cite{Stodolsky}, and still to be fully developed (see \cite{Refs} for significant 
progress in this direction): as the precision of long baseline neutrinos experiments will increase, 
the standard model will be tested in a general Riemannian geometry, extending the realm of quantum 
field theory in curved spacetimes so far developed mainly for quantum electrodynamical phenomena 
involving photons \cite{Birrell}. 

We consider the Lagrangian density for two Higgs doublet fields in a generic curved spacetime 
in which a non-minimal coupling to the Ricci scalar $R$ is added to the standard model Lagrangian as 
\begin{equation}
{\cal L}_{\mathrm{Higgs-Curvature}}=\xi_R R(\phi_1^2+\phi_2^2),
\end{equation}
where $\phi_1$ denotes the Higgs doublet of the standard model coupled to quarks and charged 
leptons, and $\phi_2$ is the Higgs doublet proposed in \cite{Wang,Gabriel}, giving mass to 
neutrinos with Yukawa couplings comparable to the ones of the charged fermions and light 
quarks. The parameter $\xi_R$ measures the coupling strength between the 
Higgs fields and the curvature scalar $R$ and, to ensure compatibility with the equivalence 
principle, we assume that $\xi_R$ has the same value for the two Higgs doublets. 

In the spontaneously broken phase, by denoting with $\mu_i$ and $\lambda_i$ the usual mass 
and self-interaction Higgs parameters (with $i=1,2$ for a two-Higgs-doublet model), the 
Higgs fields develop vacuum expectation values $v_0^{(i)}={(-\mu_{i}^2/\lambda_{i})}^{1/2}$, 
which is equal to $v_0^{(1)}=250$ GeV for the Higgs doublet of the standard model, and 
$v_0^{(2)}\simeq 10^{-2}$ eV for the Higgs doublet only coupled to neutrinos \cite{Wang,Gabriel}. 
The presence of two VEVs differing by 13 orders of magnitude originates a hierarchy problem 
quite similar, even in quantitative terms, to the one generated in GUTs without supersymmetry, 
and recently the stability of quantum corrections has been studied in detail both for models 
containing Dirac and Majorana neutrino mass terms \cite{Haba}, and models considering 
only Dirac neutrinos \cite{Morozumi}.

As usual, the masses of the elementary particles are all directly proportional to the VEVs $v_i$ 
via the Yukawa coefficients of the fermion-Higgs Lagrangian density term, $m_j=y_j v_1/\sqrt{2}$ 
for charged leptons and for quarks of flavor $j$, and $m_\nu=y_\nu v_2/\sqrt{2}$ for the three 
neutrinos mass eigenstates. Hereafter we assume that the Yukawa couplings $y_i$ are of algebraic, 
rather than of dynamical character, and under this hypothesis the mass $m_i$ will be changed only 
due to changes in the Higgs VEVs. Therefore, in a curved spacetime the effective coefficient 
of the Higgs field $\mu_i^2 \mapsto \mu_i^2+\xi_R R$, and the VEV of the Higgs fields will become 
spacetime dependent through the curvature scalar as $v_0^{(i)}=\sqrt{-(\mu_i^2+\xi_R R)/\lambda_i}$.  
If we assume a {\sl bare} neutrino mass in flat spacetime $m_{\nu}^{(0)}=
y_\nu\sqrt{-\mu_{2}^2/(2\lambda_{2})}$, in the presence of curved spacetime its value will become

\begin{equation}
m_{\nu}=\frac{y_\nu}{\sqrt{2}} \sqrt{-\frac{\mu_2^2+\xi_R R}{\lambda_2}}=
m_{\nu}^{(0)} \sqrt{1+\xi_R R \Lambda_2^2}  
\end{equation} 

\noindent
where we have introduced the reduced Compton wavelength associated to the Higgs mass 
parameter $\mu_2$ as $\Lambda_2=\hbar/(\mu_2 c)$. Assuming a Higgs mass of value 
comparable to $v_2 \simeq 10^{-2}$ eV, we obtain a Compton wavelength for the Higgs 
mass parameter $\Lambda_2 \simeq 2 \times 10^{-5}$ m. Notice that, under the hypothesis 
that the Higgs-curvature coupling strength is universal, all neutrino flavors undergo 
the same mass shift in the presence of a given gravitational source. 

We complement this analysis by also discussing the case of a coupling of the 
Higgs field to the curvature via another invariant, such as the Kretschmann 
invariant defined as $K=R_{\mu\nu\rho\sigma}R^{\mu\nu\rho\sigma}$, where $R^{\mu\nu\rho\sigma}$ 
is the curvature tensor. This invariant plays an important role in quadratic theories of 
gravity \cite{Deser,Stelle,Hehl} and has been already used in \cite{Onofrio} to 
estimate putative Higgs shifts arising from atoms in proximity of spherically 
symmetric astrophysical strong-gravity sources, and in \cite{Onofrio1} to discuss
limits coming from violations to the superposition principle for gravitational 
forces in the weak-field limit. In the weak-field limit the Higgs-curvature Lagrangian 
term will be written as
\begin{equation}
{\cal L}_{\mathrm{Higgs-Curvature}}=\xi_K \Lambda_{\mathrm{Pl}}^2 K (\phi_1^2+\phi_2^2),
\end{equation}
where $\Lambda_\mathrm{Pl}=(G\hbar/c^3)^{1/2}$ is the Planck length, whose value is 
$\Lambda_\mathrm{Pl} \simeq 10^{-35}$ m in conventional quantum gravity models or 
$\Lambda_\mathrm{Pl} \simeq 10^{-19}$ m in models with {\sl early} unifications 
of gravity to the other fundamental interactions via extra dimensions \cite{Arkani}.
By replacing the Ricci scalar coupling in Eq. (2) with the Kretschmann scalar coupling we obtain
\begin{equation}
m_{\nu}=\frac{y_\nu}{\sqrt{2}} \sqrt{-\frac{\mu_2^2+\xi_K \Lambda_{\mathrm{Pl}}^2 K}{\lambda_2}}=
m_{\nu}^{(0)} \sqrt{1+\xi_K \Lambda_{\mathrm{Pl}}^2 K \Lambda_2^2}  
\end{equation}
Since the Kretschmann invariant is nonzero outside a massive source, in this case we expect 
mass modulation even outside the mass, although with an amplitude quickly fading away 
from the source. However, due to the suppression due to the presence of the Planck length, 
the values of $\xi_K$ necessary to reproduce mass modulations of amplitude comparable to 
the one of the Ricci coupling are much larger, {\it i.e.} $\xi_K >> \xi_R$.

It may be worth to remark that under proper combinations of signs for $\xi_R$ and the Ricci 
scalar $R$, a curved spacetime with enough coupling to the Higgs field may yield a sign 
inversion in the argument of the square root in Eq. 2 and therefore an imaginary mass, 
leading to a transition from bradyonic to tachyonic behavior for neutrinos. 
Models for tachyonic neutrinos have been already envisaged several decades ago 
\cite{Mignani,Chodos,Giannetto,Hughes}, and very recently several contributions 
have appeared with a variety of mechanisms to accommodate the earlier OPERA outcome 
in a variety of scenarios (see \cite{Ma} for a review). 
While the OPERA result has been lately both retracted after identifying a systematic 
effect and shown to be incompatible with the one obtained in the contiguous ICARUS detector at LNGS 
\cite{Antonello}, it may be still premature to definitely rule out the possibility that neutrinos 
are particles endowed with a superluminal behavior \cite{Note}. Previous results as the one reported by 
the MINOS collaboration, $\delta_c=(v-c)/c=(5.1\pm 2.9) \times 10^{-5}$ (at 68 $\%$ confidence level) 
\cite{Adamson} and, at a weaker confidence level, various evidences for a negative central 
value of the muonic neutrino mass from precision studies of the pion decay 
\cite{Shrum,Backenstoss,Lu,Anderhub,Abela,Jeckelmann}, show that superluminal 
neutrino propagation is still not ruled out at a high level of confidence.
In any event, gravitationally-induced superluminality is potentially expected 
to play a role only inside or near strong-gravity sources, ruling out the possibility 
of terrestrial experiments for its verification. For instance, the Ricci scalar can 
be simply estimated inside a neutron star in the hypothesis of homogeneous mass distribution. 
For a neutron star of mass $M_n=2M\odot$ and radius $r_n=$12 Km, the Ricci scalar is 
$R_n=6 G M_n/(c^2r_n^3) \simeq 10^{-8}$ m${}^{-2}$. 
Then the superluminal regime occurs inside the neutron star if $\xi_R$ is negative and 
larger in absolute value than $|\xi_R|> 2.5 \times 10^{17}$. 
This value of nonminimal coupling is much smaller than the one assumed in \cite{Bezrukov1,Bezrukov2} 
in the framework of a proposal for the Higgs field as responsible for inflation, in which a value 
$\xi_R \simeq 10^4$ in units for which $\hbar=c=1$, corresponding to 
$\xi =10^4 (c/\hbar)^2=8 \times 10^{88}$ using MKSA units for $\hbar$ and $c$, 
was discussed. For fermions other than neutrinos, coupled to the large VEV Higgs doublet of 
the standard model, it is easy to verify that the mass shift in the presence of spacetime 
curvature inside a neutron star is negligible under any reasonable assumption unless an 
unrealistically large $\xi_R$ is assumed, since its effectiveness scales as the ratio of 
the squared Compton wavelengths of the Higgs doublets which is, in turn, proportional 
to the ratio between the VEVs of the two doublets, $v_1/v_2 \simeq 2.5 \times 10^{13}$.  

We envisage at least two frameworks in which a possible non-minimal coupling between 
the second Higgs doublet and the spacetime curvature could give rise to observable effects. 
First, since the discussed mechanism applies identically to all neutrino flavors, we expect 
that the square mass difference is also changed in the presence of a non-zero Ricci scalar, 
and this induces a modulation of the oscillation parameter for the 
$\nu_i \longleftrightarrow \nu_j$ oscillation 
$\Delta m(ij)^2=\Delta m(ij)_0^2 (1+\xi R \Lambda_2^2)$ which depends on Earth's parameters 
such as its radius and mass, unlike the oscillation parameter $\Delta m(ij)_0^2$ related 
to the neutrinos bare masses. In principle solar neutrino oscillations could also be 
affected by this effect, resulting in a contribution to the day-night measurements 
already performed \cite{Ahmad},  but it is easy to show that this is negligible 
for a broad range of nonminimal couplings due to the independence of the measurable 
average survival probability upon the source-detector distance.
Second, neutrinos created during supernovae events, depending on the location
of their production into the collapsing core, will be also affected by this
mechanism. Neutrinos generated closer to the core center will experience 
larger mass shifts with respect to the one originated in the outer layers, and 
this will imply a further spreading in the arrival time on terrestrial detectors. 
In this regard, the time lapse between neutrino events detected in different 
laboratories during SN 1987A \cite{Arnett} could be reanalyzed through this 
mechanism of mass dispersion. Furthermore, we expect a large spreading during 
the stage of abrupt truncation of the neutrino flux occurring in the early stage of 
formation of the black hole possibly resulting from the core-collapse supernova \cite{Beacom}. 

Finally, we briefly discuss possibility of modulating the mass via properly tailored structures. 
The Compton wavelength of the second Higgs doublet falls in the micrometer to hundred micrometers 
range, depending on its VEV conjectured to be in the $10^{-3}$-$10^{-1}$ eV range. 
One may therefore envisage extended structures made of many parallel layers of material 
spaced by comparable distances which should suppress the propagation of the Higgs mode 
in the space in between, analogously to the well-known inhibition of electromagnetic 
modes in a conducting cavity \cite{Purcell,Kleppner}. 
In the region occupied by this structure the neutrinos will propagate as massless or lighter particles, 
depending on the degree of cavity-induced suppression of the VEV of the second Higgs doublet.   
Interposing this structure along the path of a neutrino beam would therefore allow for the 
detection of a smaller time of flight with respect to the free propagation with full action 
of the second Higgs doublet. If the suppression factor in the electroweak cavity for antineutrinos 
differs from the one for neutrinos, a comparison between muonic neutrinos and antineutrinos beams 
should also result in different times of flight.

In conclusion, we have identified a mechanism leading to a variability of the neutrino mass 
in the presence of non-minimal coupling to curvature invariants of a light Higgs doublet 
already considered in various extensions of the standard model. 
The existence of this mechanism does not seem to clash with any experiment or observation 
available so far, stressing the importance of systematically exploring bounds on the curvature 
coupling constant to the Ricci scalar $\xi_R$ or to the Kretschmann invariant $\xi_K$. 
Possible tests of this effect have been qualitatively discussed such as atmospheric 
neutrino oscillations, the identification of a further source of spreading for 
the time of flight  originated in the SN 1987A supernova event, and the possibility 
of gravitationally-induced superluminality inside strong-gravity astrophysical objects. 
As a further test unrelated to the presence of spacetime curvature, we have proposed 
engineered electroweak cavities able to suppress the Higgs propagation, which could allow 
to modulate the time of flight of freely propagating neutrinos as in the CERN-LNGS experiments.


\begin{thebibliography}{99}
\bibitem{Note} The window of opportunity for superluminal propagation of neutrinos is 
now severely constrained through the recent result from the ICARUS collaboration in which an r.m.s. 
value of the average of the recorded distribution for the time of flight delay between photons and 
neutrinos is $\delta t=\mathrm{tof}_{\mathrm \gamma}-\mathrm{tof}_{\mathrm \nu}=(0.3 \pm 4.0 (\mathrm{stat.}) 
\pm 9.0 (\mathrm{sys.}))$ ns over an expected time of flight for the photons of $(2439098 \pm 1.5)$ns 
\cite{Antonello}.
\end{thebibliography}
\end{document}